\documentclass[english]{article}
\usepackage{graphicx,color}
\usepackage{amsmath,amssymb}
\usepackage{authblk}
\def\bc{\begin{center}}
\def\ec{\end{center}}

\def\beq{\begin{equation}}
\def\eeq{\end{equation}}

\def\bun{{\bf 1}}

\begin{document}


\title{Quantized time in quantum walks under weak rank-$K$ measurements}

\author{Klaus Ziegler$^{1,2}$\\
$^1$Institut f\"ur Physik, Universit\"at Augsburg,
D-86135 Augsburg, Germany\\
$^{2}$Physics Department, New York City College of Technology,\\
The City University of New York, Brooklyn, NY 11201, USA
}
\date{\today}

\maketitle

Measurements 
can be used to monitor the evolution of quantum systems and may lead to a universally quantized time
statistics. It is known that the mean return time is quantized for strong and indirect monitoring through the
winding number of the return amplitude in a one-dimensional space.
Here we discuss that under multi-channel strong or indirect monitoring, where the latter is achieved through 
ancilla coupling, the mean return time of a quantum walk in the projected subspace is also quantized. This 
reflects a universal time quantization for a higher dimensional evolution.

\section{Introduction}

Repeated measurements provide an efficient method to monitor the evolution of a
quantum walk~\cite{nielsen10,kempe03,qiang24}, where we distinguish between
measurements either directly by a projection~\cite{krovi06,Varbanov2008,gruenbaum13,dhar15}
or indirectly by a projection on an ancilla that is coupled to the evolving quantum 
system~\cite{gebhart20,wang22,heine25}.
Most of the recent works assume strong measurements, modelled by rank-one projectors acting 
directly on the system 
\cite{krovi06,Varbanov2008,gruenbaum13,friedman17,liu20}. 
The advantage of indirect measurements is that by tuning the coupling of the ancilla we can continuously vary 
the strength of the measurement, ranging from the uncoupled unitary evolution to the limit of strict
projective measurements. The latter results in a quantized mean return time to the initial state, which is
controlled by the winding number of the return amplitude~\cite{gruenbaum13}. In two recent papers it was
shown that the winding number $w$ of the return amplitude also plays a crucial role for the indirect measurements, 
where the mean number of measurements $\bar{n}$ to return to the initial state obeys the scaling behavior 
$\bar{n}=w/\eta$ with the ancilla coupling parameter $\eta$~\cite{heine25,ziegler26}. This was obtained under the condition that the 
initial state is pure and the projector has rank 1. Projective measurements with rank-$K$ projections have als been discussed
recently~\cite{bourgain14,liu26}. The goal of the present work is to extend to latter to indirect measurements and to extend
the indirect measurements to projections of higher dimensional spaces. In this context we will define a winding number
for a set of return and transition amplitudes and analyze the role of this winding number for the mean number of
measurements.  

The paper is organized as follows: 
In Sect. \ref{sect:return} the recurrence of the quantum walk with probability 1 on the $K$-dimensional subspace is studied,
followed by the calculation of the mean first return time and the corresponding winding number in Sect. \ref{sect:winding}. 
Then in Sect. \ref{sect:discussion} the results of this work are briefly summarized. Finally, details of the winding number
calculation (in App. \ref{app:winding}) and the normalization of the transition matrix (in App. \ref{app:norm}) are
presented.

\section{Return probability}
\label{sect:return}

We consider a projector of rank $K$ 
\beq
P=\sum_{l=1}^K|\psi_l\rangle\langle\psi_l|
,
\eeq
where $\{|\psi_l\rangle\}$ are orthogonal quantum states of a $d$-dimensional Hilbert space
with $d> K$, and calculate the return properties with indirect measurements,
extending the previous work on single-state projective and indirect measurements~\cite{heine25,ziegler26}. 

The monitored quantum walk on the projected space is characterized by the evolution matrix 
$M_n$ with matrix elements
\beq
M_{n;jj'}:=\langle\psi_j|U(QU)^{n-1}|\psi_{j'}\rangle
\ \ (1\le j,j'\le K)
,
\eeq
where $Q=\bun-x P$ and the unitary evolution operator $U=e^{-iH\tau}$. $x$ characterizes the 
coupling strength $\eta$ with the ancilla through $x=1-\sqrt{1-\eta}$. $M_{n;jj'}$ is the transition amplitude
for a quantum walk from $|\Psi_{j'}\rangle$ to $|\Psi_{j}\rangle$ during the time $n\tau$ with $n-1$
intermediate measurements, while $|M_{n;jj'}|^2$ is the corresponding transition probability.
This allows us to define
\beq
\label{return+prob}
\Gamma_n:={\rm Tr}_K(M^{}_nM^\dagger_n)=
{\rm Tr}[P U(QU)^{n-1}P(U^\dagger Q)^{n-1}U^\dagger]
\ \ \ (n\ge 1)
,
\eeq
from which we get with $\bun-Q^2=(2-x)x P$
\beq
\Gamma_n=\frac{1}{x(2-x)}{\rm Tr}[P U(QU)^{n-1}[\bun-Q^2](U^\dagger Q)^{n-1}U^\dagger]
=h_{n-1}-h_n
,
\eeq
where
\beq
h_{n}=\frac{1}{x(2-x)}{\rm Tr}[P U(QU)^{n-1}Q^2(U^\dagger Q)^{n-1}U^\dagger]
\ \ {\rm and}\ \
h_0=\frac{1}{x(2-x)}{\rm Tr}(P)=\frac{K}{x(2-x)}
.
\eeq
This implies
\beq
\sum_{n=1}^N \Gamma_n
=(h_0-h_1)+(h_1-h_2)+\cdots+(h_{N-1}-h_N)=h_0-h_N 
,
\eeq
where $\lim_{N\to\infty}h_N=0$. Thus we have eventually
\beq
\sum_{n\ge 1}\Gamma_n
=\frac{K}{x(2-x)}
,
\eeq
which means that $F_n=x(2-x)\Gamma_n/K$ defines a return probability. This
reflects the known result for the return probability if $K=1$~\cite{heine25}, while the result for $K>1$ 
indicates that  there are $K$ channels of monitored return and transition inside the $P$-projected space, 
whose contributions add up to the total probability $1$.

\section{Return amplitudes, winding numbers and quantized return times}
\label{sect:winding}

The Fourier transform of the return and transition amplitudes 
\beq
\label{fourier1}
\hat{M}(e^{i\omega})
=\sum_{n\ge 1}e^{i\omega n}M_{n}
\ \ \ (0\le \omega<2\pi)
\eeq
reads with $z=e^{i\omega}$ and $y=zx$
\beq
\hat{M}(z)=z(\langle\psi_j|(U^\dagger+yP-z)^{-1}|\psi_{j'}\rangle)
.
\eeq
From the relation $B^{-1}=A^{-1}+A^{-1}(A-B)B^{-1}$ we obtain
\beq
(U^\dagger+yP-z)^{-1}=(U^\dagger-z)^{-1}-(U^\dagger-z)^{-1}yP(U^\dagger+yP-z)^{-1}
,
\eeq
such that
$
\hat{M}=\hat{M}_0-x\hat{M}_0\hat{M}
$ with $\hat{M}_0=z(\langle\psi_j|(U^\dagger-z)^{-1}|\psi_{j'}\rangle)$,
which is equivalent to 
\beq
\label{relation1}
\hat{M}=(\bun+x\hat{M}_0)^{-1}\hat{M}_0
.
\eeq
Then we get
\beq
\int_0^{2\pi}{\rm Tr}_K(\hat{M}\hat{M}^\dagger)\frac{d\omega}{2\pi}
=\sum_{n,n'\ge 1}\int_0^{2\pi}e^{i\omega (n-n')}\frac{d\omega}{2\pi}{\rm Tr}_K(M^{}_nM^\dagger_{n'})
=\sum_{n\ge 1}{\rm Tr}_K(M^{}_nM^\dagger_{n})=\sum_{n\ge 1}\Gamma_n=\frac{K}{x(2-x)}
,
\eeq
which means that the matrix function $\hat{M}\sqrt{x(2-x)/K}$ is normalized:
\beq
\frac{x(2-x)}{2\pi K}\int_0^{2\pi}{\rm Tr}_K(\hat{M}\hat{M}^\dagger)d\omega=1
.
\eeq
Moreover, the mean number of measurements for the FDR reads
\[
\bar{n}=\sum_{n\ge 1}nF_n
=\frac{x(2-x)}{K}\sum_{n\ge 1}n{\rm Tr}_K(M^{}_nM^\dagger_{n})
\]
\beq
\label{winding}
=\frac{ix(2-x)}{2\pi K}\sum_{n,n'\ge 1}\int_0^{2\pi}e^{i\omega n}\partial_\omega e^{-i\omega n'}d\omega
{\rm Tr}_K(M^{}_nM^\dagger_{n})
=\frac{ix(2-x)}{2\pi K}\int_0^{2\pi}{\rm Tr}_K(\hat{M}\partial_\omega\hat{M}^\dagger)d\omega
.
\eeq
On the other hand, the winding number is defined as (cf. Eq. (\ref{winding2}) in App. \ref{app:winding})
\beq
\label{winding1}
w=\frac{i}{2\pi}\int_0^{2\pi}\frac{{\rm Tr}_K(\hat{M}\partial_\omega\hat{M}^\dagger)}{{\rm Tr}_K(\hat{M}\hat{M}^\dagger)}
d\omega
.
\eeq
For $x=1$ we have ${\rm Tr}_K(\hat{M}\hat{M}^\dagger)=K$ (cf. Eq. (\ref{norm2}) 
in App. \ref{app:norm}), which implies the relation $\bar{n}=w$. In general, for $0<x\le 1$ we have
according to Eq. (\ref{relation2})
\beq
\hat{M}=-\left[\hat{M}_0^\dagger \hat{M}_0^{-1}+1-x\right]^{-1}
\ {\rm and}\ \ 
\hat{M}^\dagger=-\left[[\hat{M}_0^\dagger \hat{M}_0^{-1}]^{-1}+1-x\right]^{-1}
,
\eeq
%
%
which yields for Eq. (\ref{winding})
\beq
\label{mean1}
\bar{n}
=\frac{ix(2-x)}{2\pi K}\int_0^{2\pi}{\rm Tr}_K\left(\left[\hat{M}_0^\dagger \hat{M}_0^{-1}+1-x\right]^{-1}\partial_\omega
\left[ [\hat{M}_0^\dagger \hat{M}_0^{-1}]^{-1}+1-x\right]^{-1}\right)d\omega
\eeq
The expansion in powers of $1-x$ gives 
\beq
\sum_{l_1,l_2\ge 0}(-1+x)^{l_1+l_2}
\frac{ix(2-x)}{2\pi K}\int_0^{2\pi}{\rm Tr}_K\left(
\left[\hat{M}_0^\dagger \hat{M}_0^{-1}\right]^{-l_1-1}\partial_\omega
\left[\hat{M}_0^\dagger \hat{M}_0^{-1}\right]^{l_2+1}\right)d\omega
\eeq
\beq
=\sum_{l_1,l_2\ge 0}(-1+x)^{l_1+l_2}(l_2+1)
\frac{ix(2-x)}{2\pi K}\int_0^{2\pi}{\rm Tr}_K\left(
\left[\hat{M}_0^\dagger \hat{M}_0^{-1}\right]^{-l_1-1+l_2}
\partial_\omega\left[\hat{M}_0^\dagger \hat{M}_0^{-1}\right]
\right)d\omega
.
\eeq
If $l_1\ne l_2$ the integrand is the total differential 
$
\partial_\omega{\rm Tr}_K\left(
\left[\hat{M}_0^\dagger \hat{M}_0^{-1}\right]^{-l_1+l_2}
\right)
$, whose integral vanishes due to the $2\pi$-periodicity.  Thus, we get for Eq. (\ref{mean1})
\beq
\bar{n}=
\sum_{l\ge 0}(1-x)^{2l}(l+1)\frac{ix(2-x)}{2\pi K}\int_0^{2\pi}{\rm Tr}_K\left(
\left[\hat{M}_0^\dagger \hat{M}_0^{-1}\right]^{-1}
\partial_\omega\left[\hat{M}_0^\dagger \hat{M}_0^{-1}\right]
\right)d\omega
.
\eeq
While the integral is identical with that of $x=1$ in Eq. (\ref{winding}) due to $\hat{M}=-\hat{M}_0[\hat{M}_0^\dagger]^{-1}$
in this case,  the summation with respect to $l$ gives
\beq
\sum_{l\ge 0}(1-x)^{2l}(l+1)=\frac{1}{[1-(1-x)^2]^2}
,
\eeq
such that 
\beq
\bar{n}=\frac{w(2-x)x}{[1-(1-x)^2]^2}
=\frac{w}{1-(1-x)^2}
=\frac{w}{\eta}
.
\eeq
Thus, the effect of the coupling strength $\eta$ to the ancilla on the mean FDR measurements
is a renormalization of the winding number, where the latter is obtained from $\eta=1$. 

\section{Discussion}
\label{sect:discussion}

The analysis of the monitored evolution with rank-$K$ measurements has revealed that
the dynamics inside the $K$-dimensional projected space is characterized by a recurrence with probability 1.
The average time for the first detected return is given by $\bar{t}=\tau w/\eta$ with $\bar{n}=w/\eta$
intermediate measurements, where $\tau$
is the time between measurements, $w$ is the winding number $w$ of Eq. (\ref{winding1}), 
and $\eta$ is the coupling of the ancilla to the quantum walk. These results represent the extension of
the recurrence of a single quantum state under stroboscopic rank-1 projection with strength $\eta$ 
for $K=1$ in Refs. \cite{gruenbaum13,heine25,ziegler26} to
measurements with $K>1$. The main difference is that the role of the return amplitude 
$\langle\psi_0|U(QU)^{n-1}|\psi_{0}\rangle$ is replaced by the $K\times K$ matrix
$(\langle\psi_j|U(QU)^{n-1}|\psi_{j'}\rangle)$ with $1\le j,j'\le K$. Otherwise, the universal behavior
of the mean first detected return is preserved. This reflects a remarkable universality of the monitored
evolution in the projected subspace, which is associated with the winding number of the quantum system.

\appendix

\section{Winding number}
\label{app:winding}

The winding number $w$ of a complex $2\pi$-periodic function $f(\omega)=|f|e^{i\varphi(\omega)}$ 
reads
\beq
\frac{1}{2\pi}\int_0^{2\pi}\frac{d\varphi}{d\omega}d\omega
=\frac{\varphi(2\pi)-\varphi(0)}{2\pi}
.
\eeq
Now we consider the complex number ${\rm Tr}_K[\hat{M}(e^{i\omega})\hat{M}^\dagger(e^{i\omega'})]
=|\bar{\psi}|e^{i\bar{\varphi}}$, where the phase $\bar{\varphi}$ vanishes in the limit $\omega'\to\omega$.
Then the infinitesimal phase change with respect to $\omega'$ reads
\beq
\frac{d\bar{\varphi}}{d\omega'}
=\frac{d}{d\omega'}\log\left({\rm Tr}_K[\hat{M}(e^{i\omega})\hat{M}^\dagger(e^{i\omega'})]\right)
-\frac{d}{d\omega'}\log|\bar{\psi}|
,
\eeq
which implies in the limit $\omega'\to\omega$
\beq
\lim_{\omega'\to\omega}\frac{d\bar{\varphi}}{d\omega'}
=\frac{1}{{\rm Tr}_K[\hat{M}(e^{i\omega})\hat{M}^\dagger(e^{i\omega})]}
{\rm Tr}_K[\hat{M}(e^{i\omega})\frac{d}{d\omega}\hat{M}^\dagger(e^{i\omega})]
-\frac{d}{d\omega}\log|\bar{\psi}|
.
\eeq
The logarithmic term is a total differential in $\omega$ due to
\beq
\lim_{\omega'\to\omega}\frac{d}{d\omega'}\log|\bar{\psi}|
=\lim_{\omega'\to\omega}\frac{1}{2}\frac{d}{d\omega'}\log\left(
{\rm Tr}_K[\hat{M}(e^{i\omega})\hat{M}^\dagger(e^{i\omega'})]
{\rm Tr}_K[\hat{M}(e^{i\omega'})\hat{M}^\dagger(e^{i\omega})]
\right)
\eeq
\beq
=\frac{1}{2{\rm Tr}_K[\hat{M}(e^{i\omega})\hat{M}^\dagger(e^{i\omega})]}
\left\{{\rm Tr}_K[\hat{M}(e^{i\omega})\frac{d}{d\omega}\hat{M}^\dagger(e^{i\omega})]
+{\rm Tr}_K[\frac{d}{d\omega}\hat{M}(e^{i\omega})\hat{M}^\dagger(e^{i\omega})]\right\}
\eeq
\beq
=\frac{1}{2}\frac{d}{d\omega}\log\left(
{\rm Tr}_K[\hat{M}(e^{i\omega})\hat{M}^\dagger(e^{i\omega})]
\right)
.
\eeq
Thus, the integral of the total differential of the $\log|\bar{\psi}|$ vanishes due to its $2\pi$-periodicity
such that the winding number reads
\beq
\label{winding2}
w=\frac{1}{2\pi}\int_0^{2\pi}\lim_{\omega'\to\omega}\frac{d\bar{\varphi}}{d\omega'}d\omega
=\frac{1}{2\pi}\int_0^{2\pi}
\frac{1}{{\rm Tr}_K[\hat{M}(e^{i\omega})\hat{M}^\dagger(e^{i\omega})]}
{\rm Tr}_K[\hat{M}(e^{i\omega})\frac{d}{d\omega}\hat{M}^\dagger(e^{i\omega})]
d\omega
.
\eeq

\section{Normalization}
\label{app:norm}

For $\hat{M}_0$ we get the relation
\beq
\delta_{jj'} +\hat{M}_{0;jj'}
=e^{i\omega}\langle\psi_j|e^{-i\omega}\bun+(U^\dagger-e^{i\omega})^{-1}|\psi_{j'}\rangle
=-\langle\psi_j|[(U^\dagger-e^{i\omega})^{-1}]^\dagger|\psi_{j'}\rangle
\eeq
and since $\langle\psi_j|A^\dagger|\psi_{j'}\rangle=\langle\psi_{j'}|A|\psi_{j}\rangle^*$ we obtain
\beq
\label{shift0}
\bun +\hat{M}_0=-\hat{M}_0^\dagger
.
\eeq
This enables us to write for Eq. (\ref{relation1}) with $x=1$
\beq
\hat{M}=(\bun +\hat{M}_0)^{-1}\hat{M}_0=-[\hat{M}_0^\dagger]^{-1} \hat{M}_0
\eeq
or
\beq
\hat{M}=\hat{M}_0(\bun +\hat{M}_0)^{-1}=-\hat{M}_0[\hat{M}_0^\dagger]^{-1}
,
\eeq
which yields
\beq
\label{norm2}
{\rm Tr}_K(\hat{M}\hat{M}^\dagger)={\rm Tr}_K([\hat{M}_0^\dagger]^{-1} \hat{M}^{}_0\hat{M}_0^{-1}\hat{M}_0^\dagger)=K
.
\eeq
For $0<x<1$ we get from Eq. (\ref{shift0})
\beq
\bun+x\hat{M}_0=-\hat{M}_0^\dagger-(1-x)\hat{M}_0
\eeq
to write for Eq. (\ref{relation1})
\beq
\label{relation2}
\hat{M}=(\bun+x\hat{M}_0)^{-1}\hat{M}_0
=-\left[\hat{M}_0^\dagger+(1-x)\hat{M}_0\right]^{-1}\hat{M}_0
\eeq
or
\beq
\label{relation2a}
\hat{M}=\hat{M}_0(\bun+x\hat{M}_0)^{-1}
=-\hat{M}_0\left[\hat{M}_0^\dagger+(1-x)\hat{M}_0\right]^{-1}
.
\eeq

\vskip0.5cm

\noindent
{\bf Acknowledgment:}
I am grateful to Sabine Tornow, Eli Barkai and Tim Heine for a fruitful collaboration on the indirect $K=1$ measurements.

\end{document}